\documentclass[conference]{IEEEtran}
\IEEEoverridecommandlockouts
\usepackage[letterpaper, left=0.625in, right=0.625in, bottom=1in, top=0.75in]{geometry}
\usepackage{cite}

\usepackage[utf8]{inputenc}
\usepackage{times,graphics,epsfig,amsmath,xspace,endnotes,pifont,multirow,rotating,listings,amssymb,algorithmic,color,caption,nicefrac,adjustbox,todonotes,tabularx,mathtools, algorithmic,algorithm}
\usepackage{graphicx}
\usepackage{subcaption}
\usepackage{tabularx}
\newcolumntype{L}[1]{>{\raggedright\arraybackslash}p{#1}} 
\newcolumntype{C}[1]{>{\centering\arraybackslash}p{#1}} 
\newcolumntype{R}[1]{>{\raggedleft\arraybackslash}p{#1}} %

\usepackage{amsthm}

\usepackage{amsmath}
\usepackage{amssymb}
\usepackage{accents}

\usepackage{dsfont}
\usepackage{amsmath}
\usepackage{amssymb}

\newcommand{\x} {{\bf{x}}}

\renewcommand{\u} {{\bf{u}}}

\renewcommand{\P} {{\bf{P}}}

\begin{document}
\title{Capacitated Beam Placement for Multi-beam Non-Geostationary Satellite Systems}
%\title{Virtualized Network Function Placement\\for Cellular Network Slicing}


\author{\IEEEauthorblockN{Nariman Torkzaban, Asim Zoulkarni, Anousheh Gholami, and John S. Baras}
\IEEEauthorblockA{\textit{Department of Electrical and Computer Engineering}
\textit{\& Institute for Systems Research, }\\
\textit{University of Maryland, College Park, Maryland, USA}\\
Email: \{narimant, asimz, anousheh, baras\}@umd.edu}
}
% \author{\IEEEauthorblockN{Nariman Torkzaban}
% \IEEEauthorblockA{Department of Electrical\\
% and Computer Engineering \&}
% \IEEEauthorblockA{Institute for Systems Research\\
% University of Maryland\\
% College Park, Maryland, USA\\
% narimant@umd.edu}
% \and
% \IEEEauthorblockN{Chrysa Papagianni}
% \IEEEauthorblockA{Nokia Bell Labs\\\
% Antwerp, Belgium\\
% chrysa.papagianni@nokia-bell-labs.com}
% \and
% \IEEEauthorblockN{John S. Baras}
% \IEEEauthorblockA{Department of Electrical \\
% and Computer Engineering \&}
% \IEEEauthorblockA{Institute for Systems Research\\
% University of Maryland\\
% College Park, Maryland, USA\\
% baras@isr.umd.edu}}


\maketitle

\begin{abstract}
Non-geostationary (NGSO) satellite communications systems have attracted a lot of attention both from industry and academia, over the past several years. Beam placement is among the major resource allocation problems in multi-beam NGSO systems. In this paper, we formulate the beam placement problem as a \emph{Euclidean disk cover} optimization model. We aim at minimizing the number of placed beams while satisfying the total downlink traffic demand of targeted ground terminals without exceeding the capacity of the placed beams. We present a low-complexity deterministic annealing (DA)-based algorithm to solve the NP-hard optimization model for near-optimal solutions. We further propose an extended variant of the previous model to ensure the traffic assigned to the beams is balanced. We verify the effectiveness of our proposed methods by means of numerical experiments and show that our scheme is superior to the state of the art methods in that it covers the ground users by fewer number of beams on average. 
% \asim: perhaps discuss computational complexity or say efficient instead of low-complexity?
\end{abstract}



% \fi
\IEEEpeerreviewmaketitle

\section{Introduction}
Satellite systems are envisioned to complement and extend the sixth generation of wireless communications, owing to their global coverage and variety of provided services. Due to their short round-trip delays, less stringent power consumption, and reduced launch complexity, non-geostationary (NGSO) satellite constellations have lately received much interest to be deployed on large scales. OneWeb and SpaceX are among the low earth orbit (LEO) \cite{Di19}, and O3b is among the medium earth orbit (MEO) \cite{O3b} satellite communications projects launched in the past several years.

As opposed to the conventional uniform radio resource distribution across the coverage area in traditional satellite systems \cite{Ram21}, in the evolved satellite communications systems, smart and adaptive resource allocation mechanisms employing effective beamforming techniques \cite{Wan21} are required to increase the transmission data rates. The demand for increased transmission data rates results in beamforming schemes that resort to excessive frequency reuse, which, due to the ensuing risk of inter-beam interference, complicate the design of multi-beam satellite communication architectures, that follow the cellular network architecture paradigm. 
At the same time, energy efficiency is depleted by the growing network traffic enabled by the efficient frequency reuse schemes, an issue that becomes even more challenging in the case of NGSO satellites where mobility hinders energy efficiency as well. In order for the next-generation mobile networks to address the issue of global coverage in a sustainable way, the factors of energy and spectral efficiency need to be taken into account and beamforming schemes jointly optimizing both of them, such as \cite{Wan21}, are becoming relevant.
% \nariman{asim: can you please write a paragraph here about the multi-beam satellite communications systems? with a focus on beamforming. I want to let the reader know why beamforming and multi-beam are important in NGSO satellites. Use references as appropriate. \cite{Wan21} is a good start. Have in mind that we will take a short summary of that and use it in this version for the conference and a long version of that for the journal. }

The resource allocation problem in multi-beam NGSO satellite systems is composed of multiple phases, namely, \emph{beam layout design} \cite{Mar21,Jea21,Wan22}, \emph{beam placement and beam-user mapping} \cite{Tak20, Nil21, Pac21, Din22, Bui22}, \emph{frequency assignment} \cite{Pac21}\cite{Liu21}, and \emph{power allocation} \cite{ Zin21}\cite{ Van22}. In this paper, we study the \emph{beam placement} problem in multi-beam  NSGO satellite communication systems. 
% \nariman{Anousheh: Please write a review for these 6 papers: \cite{Cam19, Tak20, Nil21, Pac21, Din22, Bui22}. A short version will be used for the conference paper and a long version for the journal. Try to have a comparison view with our paper and discuss how our paper stands out compared to these works.}
Recently, the beam placement problem has attracted significant attention in different contexts. 

The main focus of \cite{Tak20} is to optimize the placement of a fixed number of beam spots given the derived relationships between the placement of multi-spot beams and the overall system throughput. However, the proposed beam spot management method assumes a one-dimensional distribution for the demand points. Thus, the resulting spot beam centers lie in a straight line which is not practical. Moreover, in contrast to \cite{Tak20} we consider the minimization of the number of beams needed to meet the demand of ground terminals under capacity and load balancing constraints that are ignored in \cite{Tak20}. Authors in \cite{Nil21} address the beam placement and frequency assignment problems. They transform the beam placement problem to an edge clique cover problem by constructing a graph of ground users where a pair of users are connected if they can be covered by a single beam for every possible location of the satellite. Consequently, the solution to the beam placement problem becomes equivalent to finding the smallest set of maximal cliques, each assigned to a beam, that covers all users. Since the edge clique problem is known to be $\mathcal{NP}$-hard, they propose an incremental heuristic algorithm based on selecting maximal cliques with larger sizes and minimum collision (defined as the number of common nodes) until all users are covered. Our proposed beam placement scheme outperforms the heuristic algorithm of \cite{Nil21}. A multi-objective optimization formulation using a graph-based representation is presented in \cite{Pac21} and a genetic algorithm is used to solve it. In \cite{Din22}, a Quadratic Unconstrained Binary Optimization (QUBO) formulation is proposed to solve the beam placement problem by using quantum computing. 

In contrast to the proposed methodologies of \cite{Nil21}, \cite{Pac21}, and \cite{Din22}, we take into account the terminals' traffic demands, satellites beam capacity constraint and load balancing constraint. 
%\anoushe{in the evaluation part, \cite{Pac21} considers an independent freq assignemnt on top of their beam placement algo, and presents other metrics like power. we could do the same in future.}
In \cite{Liu21}, the beam resource allocation and scheduling problems are addressed. In order to improve the overall quality of service and due to the gain attenuation at the edge of the beams compared to the center, an incremental beam placement procedure is proposed where the satellite beams are placed such that their centers are aligned with the location of the users with high demands and the remaining users are covered by adding new beams until all users are covered. %we could compare DAto this method too

In this paper, we propose a capacity-aware beam placement algorithm 
 to serve the downlink traffic of a group of target mobile users, while at the same time, ensuring that the traffic load allocated to the employed beams is balanced. We note that given the high overhead of adaptive beamforming schemes, custom beam footprint design techniques cannot be practically adopted in fast-paced NGSO multi-beam satellite systems. Hence, similar to \cite{Pac21,Bui22}, we opt for placing conical beams with circular footprints on the earth's surface. The main contributions of this paper are as follows.
 \begin{itemize}
     \item We formulate beam placement as an instance of the \emph{Euclidean disk cover} problem with the goal of minimizing the number of placed beams, while ensuring the traffic assigned to each beam does not exceed the beam capacity. 
     \item Given the disk cover problem is $\mathcal{NP}$-hard, based on the notion of the $H^{\infty}$ norm, we propose an approximation model for the beam placement problem. Inspired by \cite{Per08}, we propose a low-complexity \emph{deterministic annealing} (DA)-based algorithm to solve the approximate model for near-optimal solutions.
     \item We refine our solution from the last step by augmenting our optimization model to ensure the traffic load assigned to the placed beams is balanced. 
     \item Finally, we validate the effectiveness of our algorithm by means of numerical experiments and benchmarking against conventional methods. 
 \end{itemize}
     
 The remainder of the paper is organized as follows. Section~\ref{sec:desc} describes the network model and the problem description. In Section~\ref{sec:problem} we formulate the beam placement problem and obtain the approximate model. We propose our DA-based algorithm for beam placement and its refinement for load balancing in section~\ref{sec:proposed}, while section \ref{sec:evaluation} presents our evaluation results. Finally, we highlight our conclusions in Section \ref{sec:conclusions}.

\section{Problem Definition} 
\label{sec:desc}
\subsection{System Model}
\begin{figure}
\centering
\includegraphics[width=0.47\textwidth]{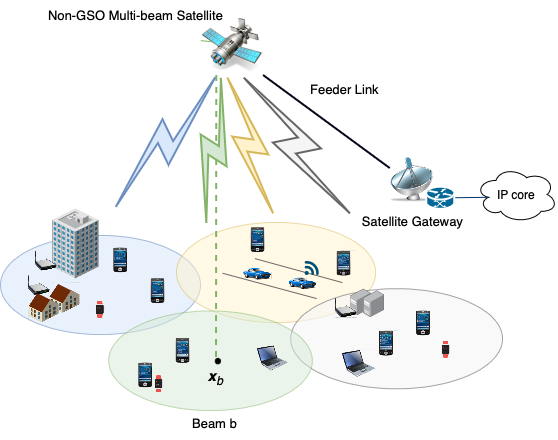}
  \caption{NGSO Multi-beam Satellite-aided System model } % \asim: is it possible to have an svg or pdf format of that image?
    \label{fig:model}
\end{figure}

We consider the downlink of an NGSO multi-beam integrated satellite-terrestrial network as depicted in Fig. \ref{fig:model}. The NGSO satellite employs a multi-antenna array and is therefore enabled to generate a set of space-ground beams. The satellite gateway is connected to the space segment by means of the \emph{feeder link}. Any ground terminal that can communicate with the satellite segment is considered a ground \emph{user}. The downlink happens via the \emph{user links} that are realized through the space-ground beams. Let the ground segment consist of $U$ ground users denoted by the set $\mathcal{U}$, where each user $u\in\mathcal{U}$ is located on the earth's surface at the coordinates $(x_u^{(1)}, x_u^{(2)})$. The ground segment is to be served through beams of an NGSO multi-beam satellite that is capable of generating the transmission beam set $\mathcal{B}$ consisting of $B$ beams. Similar to \cite{Nil21}, we consider conical beams with a circular cross-section on the earth's surface. Let $\x_b = (x_b^{(1)}, x_b^{(2)})$ denote the center of the beam $b\in \mathcal{B}$. Further, let $r_{max}$ and $c_{max}$ denote the radius of the cross-section and the capacity of the space-ground links associated with each beam.

\subsection{Problem Description}
The \emph{beam placement} problem in the simplest form entails finding the minimum number of beams $B$, and the optimal locations of their centers $x_b, b\in[B]$, that can collectively \emph{cover} a target group of ground users $\mathcal{U}$ while satisfying the traffic demand of all such users. The user set $\mathcal{U}$ is covered by the beam set $\mathcal{B}$ if every node $u\in\mathcal{U}$ is covered by \emph{exactly} one beam from the set ${B}$. In other words, the solution to the beam placement problem will also reveal an \emph{assignment policy} matrix $\P \in \mathbb{Z}^{B\times U}$ where $p_{b|u} =1$ if the ground node $u \in \mathcal{U}$ is served by beam $b\in \mathcal{B}$ and $p_{b|u} =0$ otherwise. 

A couple of notes are in order; First, for energy efficiency concerns constrained by the active placed beams, fractional assignment of the traffic of ground terminals to the placed beams is not allowed in the beam placement solution, i.e. each user must be served by only one beam. Second, various objectives may be pursued by solving the beam placement problem. For instance, the above problem definition ensures that all the ground users are connected to the space segment through the placed beams and therefore is denoted as the \emph{coverage} problem. This problem is already formulated in \cite{Nil21}. However, in a realistic scenario, it is of paramount importance to guarantee that the assigned traffic to each beam does not exceed the throughput that the beam can support. This requires to define a new variant of the beam placement problem which we denote by the \emph{capacity} problem. Finally, we note that the non-constrained assignment of the ground users to the placed beams may result in situations where heavy burden is enforced on some beams, while the other beams are rarely used. Therefore, it is important to emphasize on \emph{load balancing} as an important objective in the satellite beam placement problem. Such an objective is sought by the research in \cite{Bui22}. 

In the next section, we present a realistic formulation for the beam placement problem that takes into account the constraints and objectives mentioned above. 

% We consider two problems, \emph{connectivity} and \emph{capacity}.

% The mapping of the beams to the satellites is out of the scope of this paper. 

\section{Problem Formulation}
\label{sec:problem}

In this section, we formulate the beam placement problem. We note that each ground user can be covered by a beam $b\in \mathcal{B}$ only if it is within the radius $r_{max}$ of the beam center $\x_b$. Also, the total user traffic demand supported by each beam $b \in\mathcal{B}$ must not exceed the capacity of that beam $c_{max}$. Therefore, the \emph{coverage} beam placement problem is formulated  as follows. 
 \begin{align}
 \text{Minimize} \quad &M; \quad \text{subject to} \nonumber\\
& \exists \quad \P, \x_1 \ldots \x_M;\text{ } \max_{u\in\mathcal{U}}||\u-\x_{b(u)}||\leq r_{max}\label{cons:max}\\
% &\sum_{u\in\mathcal{U}} f_up_{b|u}\leq c_{max}, \quad \forall b\in \mathcal{B} \label{cons:cap}\\
&\sum_{b\in\mathcal{B}}{p_{b|u}} = 1, \quad \forall u\in\mathcal{U}\label{cons:feas}\\
& \x_m \in \mathbb{R}^2, \quad \forall m \in [M]\label{cons:dom1}\\
& p_{b|u} \in \{0,1\}, \quad \forall b \in \mathcal{B}, \quad \forall u \in \mathcal{U}\label{cons:dom2}
\end{align}
where $b(u)$ is the beam to which the user $u\in\mathcal{U}$ is assigned, $f_u$ is the traffic of the user $u\in \mathcal{U}$, and $||.||$ is the $\ell^2$-norm operator. Constraint~\eqref{cons:max} enforces the placement policy to be valid. 
% Constraint~\ref{cons:cap} ensures that the total traffic assigned to each beam does not exceed the capacity of that beam, while 
Constraint~\eqref{cons:feas} ensures that each user is assigned to only one beam (although this is implicitly assumed in the definition of the $b(u)$ notation). Constraints~\eqref{cons:dom1} and \eqref{cons:dom2} are the domain constraints. 
To formulate the \emph{capacity problem} we need to add the following capacity constraint
\begin{align}
    &\sum_{u\in\mathcal{U}} f_up_{b|u}\leq c_{max}, \quad \forall b\in \mathcal{B} \label{cons:cap}
\end{align}
 which ensures that the total traffic assigned to each beam does not exceed the capacity of that beam. The above problem is an instance of the Euclidean \emph{disk cover} problem that is known to be $\mathcal{NP}$-hard \cite{fowler_optimal_1981}. Hence, we propose an approximation method to generate near-optimal solutions with low time complexity.

% \nariman{Remember to add the constraint that ensures each user is assigned to exactly one beam. Remember to explain why we prefer this over the fractional assignment.}

\subsection{Approximate Model}
For any positive values $s_m, m=1\ldots M$, and for large enough $\alpha$, it holds that 
\begin{align}
    \max \left(s_1, \ldots, s_M\right) \cong\left(s_1^\alpha+\ldots+s_M^\alpha\right)^{\frac{1}{\alpha}} \label{eq:approx}
\end{align}

Therefore, we can approximate constraint~\eqref{cons:max} by a summation according to equation~\eqref{eq:approx}, to get
\begin{align}
    \exists \quad \x_1, \ldots, \x_M ; \quad \sum_{u\in\mathcal{U}} ||\u-\x_{b(u)}||^\alpha \leq r_{max}^\alpha \label{cons:summ}
\end{align}
and get the approximate optimization model as follows, 
 \begin{align}
 & \text{Minimize} \quad M \nonumber\\
& \text{subject to } \eqref{cons:summ}, \eqref{cons:cap}, \eqref{cons:feas}, \eqref{cons:dom1}, \eqref{cons:dom2}\nonumber
\end{align}
that is a non-convex constrained clustering problem. We use DA \cite{DA-rose} to approximately solve the last model for globally near-optimal solutions.

\section{Proposed Solution}
\label{sec:proposed}

\subsection{Deterministic Annealing}

DA is a clustering approach for assigning a large set of data points $\mathcal{X}$ to a small set of centers $\mathcal{Y}$ through minimizing an average distortion function $D=\sum_{x\in\mathcal{X}} p(x) d(x, y(x))$ where $p(x)$ is the probability of data point $x$. The DA scheme avoids local minima by converting the hard clustering problem into soft clustering where every data point $x\in\mathcal{X}$ may be assigned to multiple centers $y\in \mathcal{Y}$ through \emph{association fractions} $p(y|x)$. The distortion function can be restated as $D=\sum_{x\in \mathcal{X}} p(x) \sum_{y\in\mathcal{Y}} p(y|x) d(x, y)$. The DA scheme aims to minimize the distortion function at different randomness levels controlled by the entropy function $H(X, Y)$. More precisely, DA minimizes the objective function $F=D-T H(Y | X)$ at different temperature levels $T$ starting from high temperature and gradually decreasing the value of $T$. 

\subsection{DA-based Beam Placement Approach}

We denote by $d_{\alpha}(u, b) = ||\u-\x_b||^{\alpha}$ the distortion as a result of assigning the user $u\in\mathcal{U}$ to the beam $b\in\mathcal{B}$. We propose an iterative approach for solving the approximate optimization model. Our approach entails gradually adding and placing the space-ground beams, one at a time. At the $m$-th step, we aim at optimizing the locations of the $m$ added beams such that the left-hand side (LHS) of \eqref{cons:summ} is minimized until constraint~\eqref{cons:summ} is satisfied. We stop adding the beams when the capacity constraint~\eqref{cons:cap} is satisfied. We can express the overall distortion function as follows, 
\begin{align}
    D=\sum_{u\in\mathcal{U}} p\left(u\right) \sum_{b\in\mathcal{B}} p\left(b | u\right)\left[d_\alpha\left(u, b\right)\right]\label{eq: obj_func}
\end{align}

The DA algorithm aims at minimizing the objective function $F=D-T H(\mathcal{B} | \mathcal{U})$, where
\begin{align}
    H(\mathcal{B} | \mathcal{U})=-\sum_{u\in\mathcal{U}} p\left(u\right) \sum_{b\in\mathcal{B}} p\left(b |u\right) \log p\left(b |u\right)\label{eq:entropy}
\end{align}
and the minimization is done with respect to association fractions $p\left(b |u\right)$ where it must hold as well for all $b\in\mathcal{B}$ that $p\left(b\right)=\sum_{u\in\mathcal{U}} p\left(u\right) p\left(b |u\right)$ and for all $u\in\mathcal{U}$ that $\sum_{b\in\mathcal{B}} p\left(b |u\right)=1$ to ensure valid probability distributions are used in \eqref{eq:entropy}. The solution to this optimization problem is given by the Gibbs distribution as follows, 
\begin{align}
    p\left(b | u\right)=\frac{\exp \left(-\frac{d_\alpha\left(u, b\right)}{T}\right)}{Z_u}, \quad b\in\mathcal{B}, u \in \mathcal{U}.\label{eq:Gibbs}
\end{align}
where $Z_u = \sum_{u\in\mathcal{U}}\exp \left(-\frac{d_\alpha\left(u, b\right)}{T}\right)$. The optimal value for $F$ can be easily found as
\begin{align}
    F^*=-T \sum_{u\in\mathcal{U}} p\left(u\right) \log Z_{u}
\end{align}
that is a function of the locations of the centers of the space-ground beams. To find the optimal locations of the beam centers for minimizing $F^*$, we take the derivative of the objective function $F^*$ with respect to $\x_b = (x^{(1)}_b, x^{(2)}_b), b\in \mathcal{B}$ and set to zero to obtain
\begin{align}
    & x^{(i)}_{b}=\frac{ \sum_{u\in\mathcal{U}} u^{(i)} p\left(u\right) p\left(b | u\right) d_\alpha\left(u, b\right)^{1-2 / \alpha}}{ \sum_{u\in\mathcal{U}} p\left(u\right) p\left(b|u\right) d_\alpha\left(u, b\right)^{1-2 / \alpha}}, \text i\in\{1,2\}
\end{align}

Our approximation algorithm starts by setting $M=1$ at an initial temperature $T_{high}$. At that temperature, the unconstrained clustering problem is solved, i.e. the capacity constraint is not considered. The temperature is then reduced slowly until every ground node is covered by at least one space-ground beam and the capacity constraint is satisfied. At each step (i.e., fixed temperature) the association fractions, i.e. $p(b|u)$ are computed which allow for stating the DA objective as an explicit function of $\x_b, b\in\mathcal{B}$. Next, the optimal locations for the centers of the beams are computed. This process will continue until convergence occurs. If after a fixed number of iterations, the capacity constraints are not satisfied or there exist ground nodes that are not covered by a single ground-space beam, a new beam will be added. The process of adding the new beam is as follows; The beam center with the furthest assigned ground users or with the highest assigned traffic is chosen to spawn a new center at the same location with a small perturbation. Next, the assigned traffic to the old center is divided equally between the two centers. By our optimization mechanism, if the newest center is indeed required, it starts deviating from its old counterpart in the subsequent iterations, otherwise, if the two centers are within a predefined threshold after a few steps, they will be merged again. In the next subsection, we propose an extension of our model to address load balancing. 

\subsection{Load Balancing Extension}

To account for load balancing, we add a new term to the objective function~\eqref{eq: obj_func} to obtain the new distortion function 
\begin{align}
        D_{\text{LB}}=\sum_{u\in\mathcal{U}} p\left(u\right) \sum_{b\in\mathcal{B}} p\left(b | u\right)\left[d_\alpha\left(u, b\right) + \eta d_\beta\left( p(b)\right)\right]\label{eq: obj_func_lb}
\end{align}
where $d_\beta(p) = 1/p^\beta$ for $\beta>0$ and $\eta$ is a parameter that adjusts the balance between the two distortion terms. For a large value of $\beta$ and fractional values of $p(b)$, the second term of $D_{\text{LB}}$ blows up and the final solution moves toward a balanced load for the placed beams. Under load balancing, the DA algorithm minimizes the new objective function $F_{\text{LB}}=D_{\text{LB}}-T H(\mathcal{B} | \mathcal{U})$ to obtain for $b\in\mathcal{B}, u \in \mathcal{U}$ that,  
\begin{align}
    p\left(b | u\right)=\frac{\exp \left(-\frac{d_\alpha\left(u, b\right)+ \eta d_\beta\left(p\left(b\right)\right)+ \eta p\left(b\right) \frac{\partial d_\beta\left(p\left(b\right)\right)}{\partial p\left(b\right)}}{T}\right)}{Z^{\text{LB}}_{u}}\label{eq:Gibbs_lb}
\end{align}
 where $Z^{\text{LB}}_u = \sum_{u\in\mathcal{U}}\exp \left(-\frac{d_\alpha\left(u, b\right)+ \eta d_\beta\left(p\left(b\right)\right)+ \eta p\left(b\right) \frac{\partial d_{\beta}\left(p\left(b\right)\right)}{\partial p\left(b\right)}}{T}\right)$. We would then have for the optimal value of $F_{\text{LB}}$ that, 
\begin{align}
    F_{\text{LB}}^*=-T \sum_{u\in\mathcal{U}} p\left(u\right) \log Z_{u} - \eta \sum_{b\in\mathcal{B}}{p^2\left(b\right) \frac{\partial d_\beta\left(p\left(b\right)\right)}{\partial p\left(b\right)}} 
\end{align}

\section{Performance Evaluation}
\label{sec:evaluation}

\begin{figure}
\centering
\includegraphics[width=0.48\textwidth]{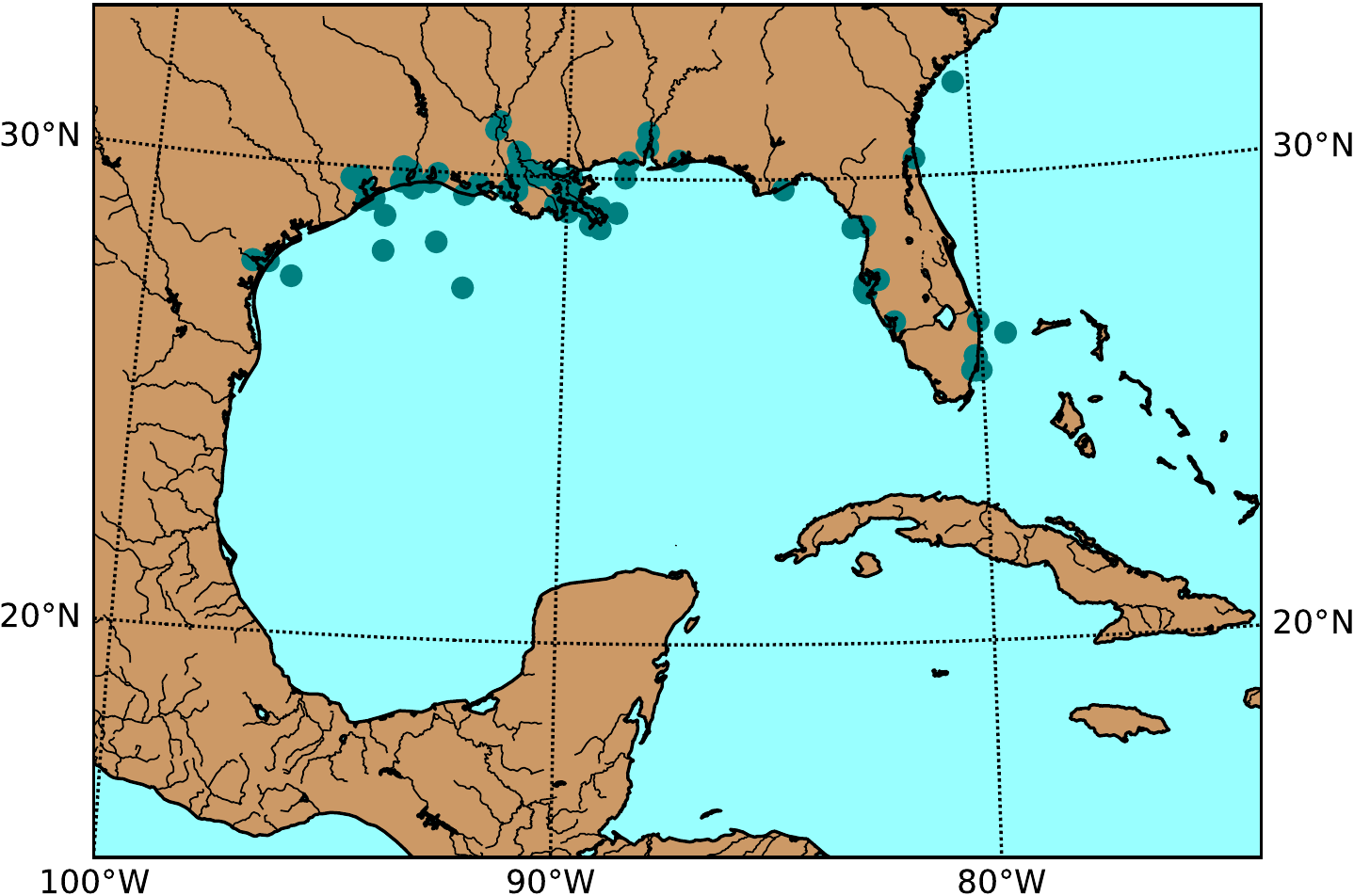}
  \caption{Locations of 100 U.S. vessels in the approximate Gulf of Mexico area, at 13:04 UTC of June 30th, 2022 }
    \label{fig:vessels}
\end{figure}
% \anousheh{baseline: grid search, K-means(ignoring capacity and load constraints)}\\
% \anousheh{comparison metric: number of beams, maximum load}

In this section, we evaluate the performance of our beam placement scheme. We first describe the simulation setup and parameters and then proceed with the numerical results. 
\subsection{Setup \& Parameters}
We consider a Starlink LEO satellite with an altitude of $1110$ km and a latitude and longitude of $26.812309^{\circ}$ and $-85.386382^{\circ}$, respectively. We assume the satellite has access to $8$ downlink channels and the frequency assignment plan allows for a frequency reuse factor of $4$. Therefore, a maximum of $32$ space-ground beams can be placed to serve the group of ground terminals. Further, we assume each satellite can deliver a throughput of $23$ Gbps \cite{Ini19} leading to a maximum capacity of $700$ Mbps per beam. Each beam has a full-cone aperture of $4.6^{\circ}$ leading to at most a $45$ km beam radius. To model the ground terminals, we extract random locations from the actual vessel locations dataset provided by the US Coast Guard \cite{marinecadastre}. Depending on the target size we sample unique data points in proximity to the Gulf of Mexico, on June 30th, 2022. An example of our test set is provided in Fig.~\ref{fig:vessels}. We set the downlink traffic rate at each location uniformly distributed in the range of $[5,10] $ Mbps. Throughout the experiments, we set $\alpha = \beta = 10$, and $\eta = 0.5$.  Our tests run on an Intel i9 CPU at 2.3 GHz and 16 GBs of main memory.

% We benchmark the performance of our proposed beam placement solution based on DA against the algorithm of \cite{Nil21} referred to as CC. We consider the following metrics for the performance comparison:
% \begin{itemize}
%     \item Number of placed beams,
%     \item Fractional beam traffic load defined as 
% \end{itemize}
% $$\frac{\sum_u p(b|u) f_u}{\sum_u f_u}$$
% Note that in the final solution of the DA algorithm, $p(b|u) = 0$ or $1 \ \forall u \in \mathcal{U}$. 

\begin{figure*}[!t]
     \centering
     % \begin{subfigure}[b]{0.24\textwidth}
     %     \centering
     %     \includegraphics[width=\textwidth]{figures/beam_placement_vis_8.36-crop.pdf}
     %     \caption{CC Beam Placement Method: 11 beams}
     %     \label{fig:obs2-1}
     % \end{subfigure}
     \begin{subfigure}[b]{0.49\textwidth}
         \centering
         \includegraphics[width=\textwidth]{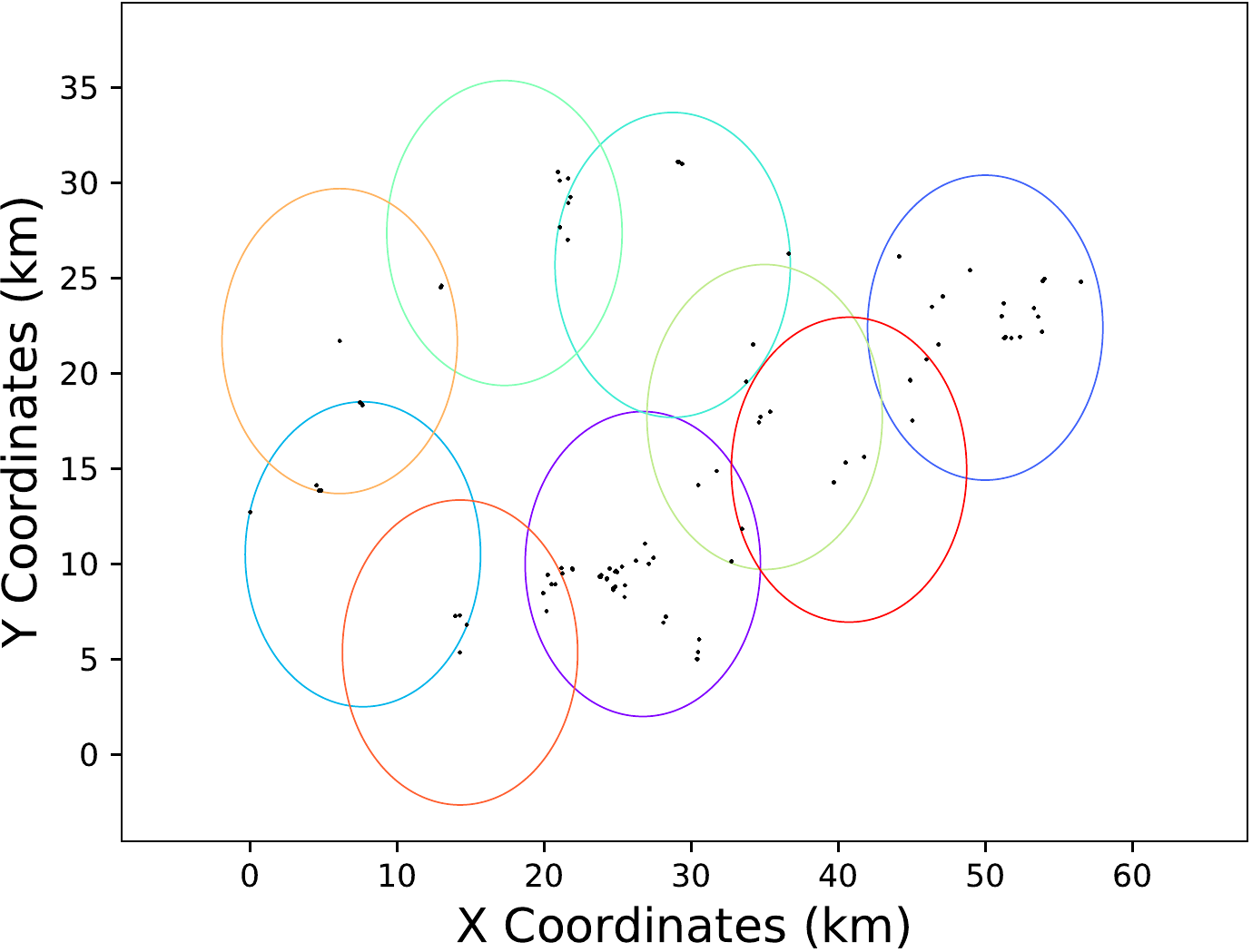}
         \caption{\emph{CC }Beam Placement Method: 9 beams}
         \label{fig:cccov}
     \end{subfigure}
     % \begin{subfigure}[b]{0.32\textwidth}
     %     \centering
     %     \includegraphics[width=\textwidth]{figures/beam_placement_vis_10.2-crop.pdf}
     %     \caption{CC Beam Placement Method: 11 beams}
     %     \label{fig:obs2-1}
     % \end{subfigure}
     \begin{subfigure}[b]{0.49\textwidth}
         \centering
         \includegraphics[width=\textwidth]{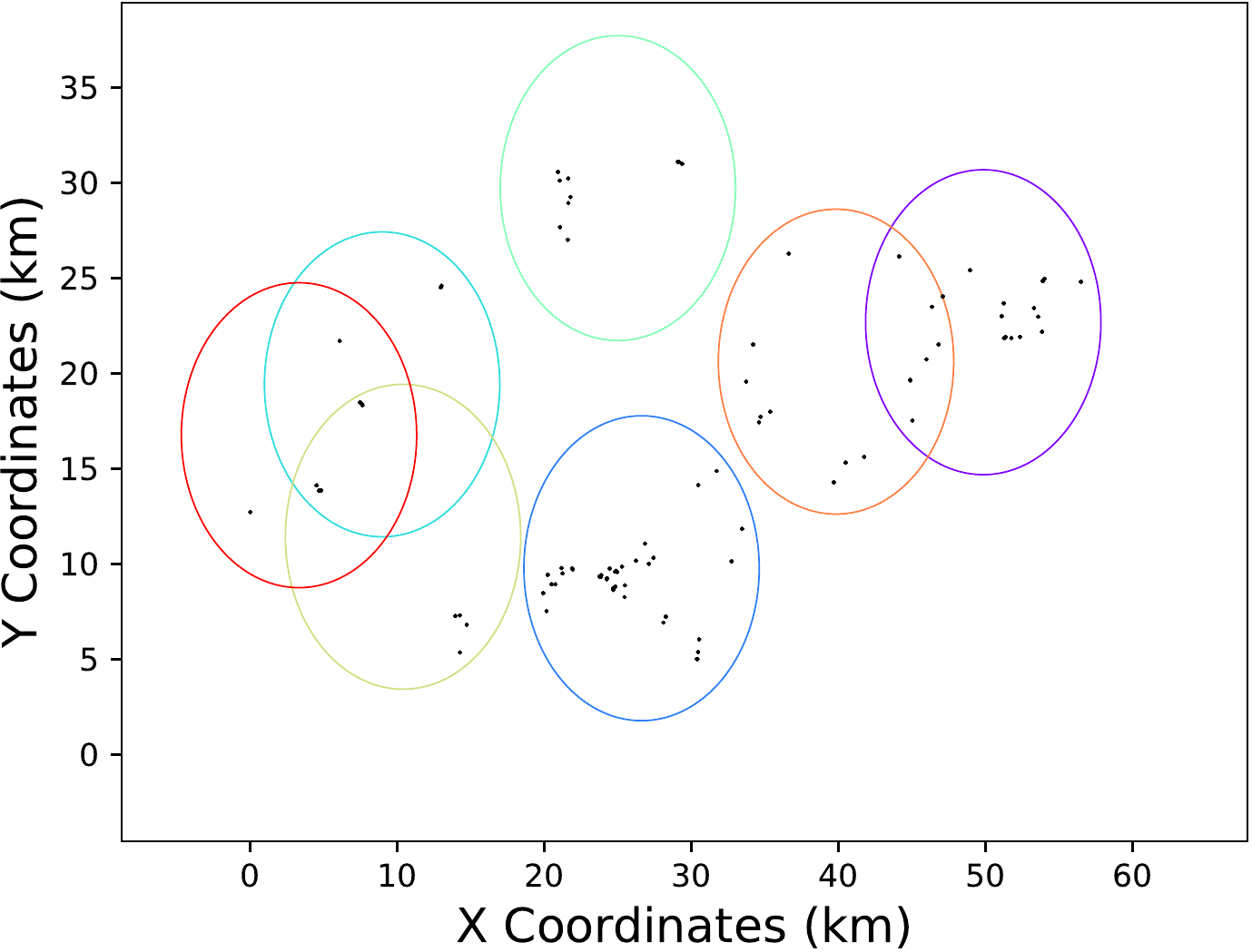}
         \caption{\emph{DA-COV }Beam Placement Method: 7 beams}
         \label{fig:dacov}
     \end{subfigure}
        \caption{Beam Placement Solution Example for the \emph{Coverage} Problem, $r_{max}=8km, U =100$: \emph{DA-COV }vs. \emph{CC }\cite{Nil21}}
        \label{fig:comp_cc_da}
\end{figure*}

\subsection{Numerical Results}
We present numerical results to validate the effectiveness of our beam placement approach for any of the \emph{coverage}, \emph{load balancing}, and  \emph{capacity} problems discussed in Section \ref{sec:problem}. For the \emph{coverage} problem, we benchmark the performance of our DA-based method against the method proposed in \cite{Nil21} that models beam placement for coverage as a clique cover problem. In \cite{Nil21}, an undirected graph is constructed by drawing edges between any two nodes that can be covered by the same beam. Then, a randomized method is presented to solve the edge clique cover problem. For the accurate implementation of \cite{Nil21}, we connect any two nodes representing vessels within distance $2\times r_{max}$ with an edge to form a similar graph and then apply the algorithm in \cite{Nil21}. Finally, we find and plot the corresponding beams with radius $r_{max}$ that cover the obtained cliques. Throughout the experiments, we denote the method in \cite{Nil21} by \emph{CC} and our proposed scheme for coverage by \emph{DA-COV}. Fig.~\ref{fig:comp_cc_da} depicts an example of beam placement result for $r_{max} = 8 Km$ where $100$ vessel locations are sampled randomly, using the \emph{CC} and \emph{DA-COV} methods, respectively. It is observed that the \emph{DA-COV} method results in capable of covering the ground users y a lower number of beams compared to the \emph{CC} method. 
% In this section, the numerical results are presented for the following three experiments corresponding to the \emph{coverage}, \emph{load balancing}, and  \emph{capacity} problems discussed in Section \ref{sec:problem}.
In the sequel, we will elaborate on the performance of our beam placement method for the \emph{coverage}, \emph{load balancing}, and \emph{capacity} problems, denoted by  \emph{DA-COV}, \emph{DA-LB}, and \emph{DA-CAP}, respectively. 

\subsubsection{Coverage Problem}

\begin{figure}[!t]
\centering
\includegraphics[width=0.48\textwidth]{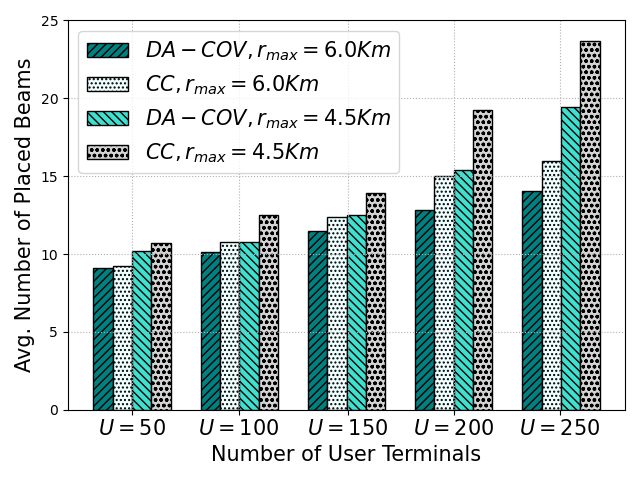}
  \caption{ \emph{DA-COV} Performance for the \emph{Coverage} Problem} % \asim: is it possible to have an svg or pdf format of that image?
    \label{fig:beamnum}
\end{figure}

Fig. \ref{fig:beamnum}, shows the average number of placed beams in the \emph{DA-COV }and \emph{CC} algorithms where the number of users, $U$ vary from $50$ to $250$ with a step size of $50$ and for two different $r_{max}$ values of $4.5Km$ and $6Km$. The number of placed beams are averaged over $10$ runs, each corresponding to an experiment with a random selection of $U$ vessel locations. It is expected that as $U$ increases and $r_{max}$ decreases, more beams will be required to cover the ground network. This can be observed for both \emph{CC }and \emph{DA-COV }algorithms as shown in Fig. \ref{fig:beamnum}. Moreover, it is observed that for any number of users and any beam radius, our proposed DA algorithm outperforms the CC scheme in terms of the average number of placed beams. We note that given the randomized nature of the \emph{CC }algorithm, various solutions may appear for different runs of the algorithm on the same test set. Therefore, for each instance, we run \emph{CC }for $100$ times and report the best solution. 

% \nariman{Plots: 1) Comparing \#beams placed in ourself with Pac21 2) comparing the time in ourself with pac21 3) In ourself, size of the beam vs. \#beams 4) In ourself, size of the beam vs. time complexity
% 5) In ourself, load balancing comparison with non-balance for a given s and eta}

\subsubsection{Load Balancing Problem}
\begin{figure}
\centering
\includegraphics[width=0.47\textwidth]{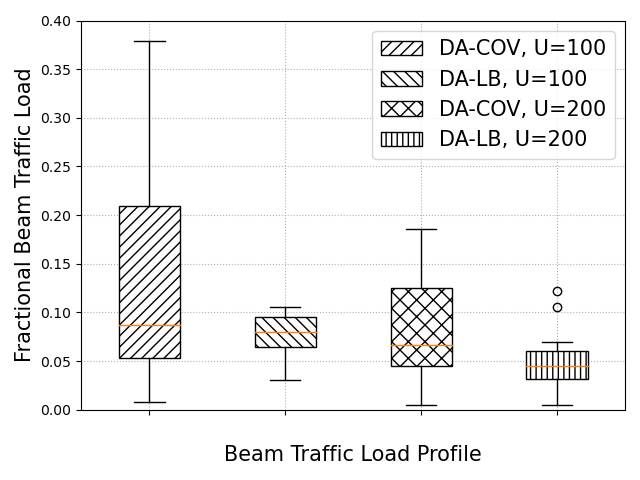}
  \caption{ \text{\emph{DA-LB} Performance for the \emph{Load Balancing}} Problem} % \asim: is it possible to have an svg or pdf format of that image? %\nariman: Yes, it will take longer to load the editor with that format.  usually convert at the end.
    \label{fig:beamloadbalance}
\end{figure}

 Figure \ref{fig:beamloadbalance} depicts the beam traffic profile for $U=100$ and $U=200$. We observe that the variation in the fractional traffic load for \emph{DA-LB }is less than \emph{DA-COV }as expected in both cases of $U=100$ and $U=200$. This is achieved by placing more beams, as the number of placed beams for \emph{DA-COV }and \emph{DA-LB }are respectively $7$ and $9$ for $U=100$. Similarly, $10$ and $12$ beams are placed by DA and \emph{DA-LB }solutions for $U=200$. Therefore, our proposed \emph{DA-LB }solution successfully addresses the \emph{load balancing} constraint in the satellite beam placement problem.

% \nariman{number of beams: 7 9 10 12, rb = 9km.}

\subsubsection{Capacity Problem}
\begin{figure}
\centering
\includegraphics[width=0.47\textwidth]{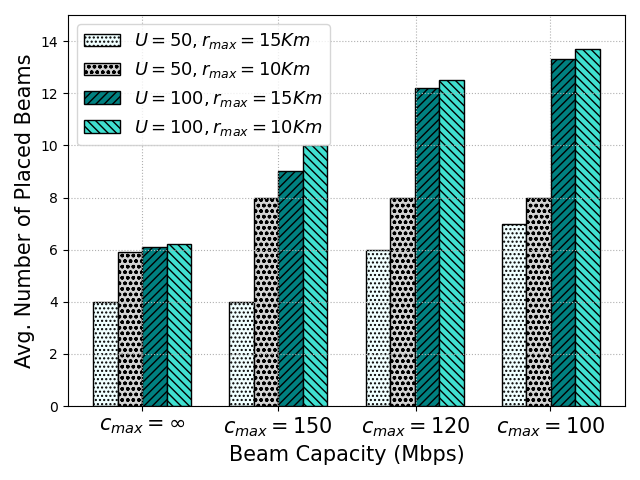}
  \caption{ \emph{DA-CAP} Performance for the \emph{Capacity} Problem} % \asim: is it possible to have an svg or pdf format of that image?
    \label{fig:beamnum-cap}
\end{figure}

We change the value of $c_{max}$ from infinity (i.e. \emph{DA-COV }problem) to $150, 120$, and $100Mbps$ and observe the average number of placed beams over $10$ runs. We also consider the cases of $U=50$ and $U=100$ with $r_{max} = 10Km$ and $r_{max} = 15Km$. Fig. \ref{fig:beamnum-cap} shows that for each pair of values for $U$ and $r_{max}$, as the beam capacity decreases, the number of placed beams increases. This is expected since the capacity degradation is compensated for by employing a higher number of satellite beams in order to fully support the ground traffic demand. This experiment highlights the importance of considering the \emph{capacity} constraints in the beam placement problem.

\section{Conclusions}
\label{sec:conclusions}
In this paper, we studied the beam placement problem in the multi-beam NGSO systems. We considered beams with circular footprints and modeled the problem as a variant of the disc cover problem. We considered three variants of the beam placement problem, namely, \emph{coverage}, \emph{capacity}, and \emph{load balancing}. We proposed an approximation scheme based on \emph{deterministic annealing} to find near-optimal solutions to the mentioned problems. Extensive numerical experiments verified that our beam placement approach covers the ground users with fewer number of beams in comparison to the state-of-the-art methods. 
% \section*{Acknowledgments}\noindent

% \nariman{Remember to update this}

% This material is based upon work supported by the Office of Naval Research award # N000141712622, Defense Advanced Research Projects Agency (DARPA), and  Leidos to the University of Maryland College Park.
\bibliographystyle{IEEEtran}
\bibliography{bibliography}
\end{document}